\documentclass[aps,prl,twocolumn,superscriptaddress,showpacs,nofootinbib]{revtex4-1}

\usepackage{graphicx,amsmath,amssymb,bm,color}

\DeclareGraphicsExtensions{ps,eps,ps.gz,frh.eps,ml.eps}
\usepackage[english]{babel}
\usepackage{fmtcount} 
\usepackage{graphicx}
\usepackage{amsmath, amsfonts, amssymb, amsthm, amscd, amsbsy, amstext} 
\usepackage{marvosym} 
\usepackage{braket,xspace}
\usepackage{mathrsfs} 
\usepackage{booktabs}
\usepackage{upgreek}
\usepackage{array}
\usepackage[makeroom]{cancel}
\usepackage{ wasysym }

\usepackage{blindtext}
\usepackage{url}
\usepackage{hyperref}%

\makeatletter
\newcommand{\overleftrightsmallarrow}{\mathpalette{\overarrowsmall@\leftrightarrowfill@}}
\newcommand{\overrightsmallarrow}{\mathpalette{\overarrowsmall@\rightarrowfill@}}
\newcommand{\overleftsmallarrow}{\mathpalette{\overarrowsmall@\leftarrowfill@}}
\newcommand{\overarrowsmall@}[3]{%
  \vbox{%
    \ialign{%
      ##\crcr
      #1{\smaller@style{#2}}\crcr
      \noalign{\nointerlineskip}%
      $\m@th\hfil#2#3\hfil$\crcr
    }%
  }%
}
\def\smaller@style#1{%
  \ifx#1\displaystyle\scriptstyle\else
    \ifx#1\textstyle\scriptstyle\else
      \scriptscriptstyle
    \fi
  \fi
}
\makeatother
\newcommand{\te}[1]{\overleftrightsmallarrow{#1}}
\newcommand{\tel}[1]{\overleftsmallarrow{#1}}
\newcommand{\ter}[1]{\overrightsmallarrow{#1}}
\newcommand{\breq}{\nonumber \\}  
\newcommand{\bs}{\mathbf}
\newcommand{\kF}{k_{\rm F}}

\def\XXint#1#2#3{{\setbox0=\hbox{$#1{#2#3}{\int}$ }
\vcenter{\hbox{$#2#3$ }}\kern-.6\wd0}}

\makeatletter

\newcommand{\Rmnum}[1]{\expandafter\@slowromancap\romannumeral #1@}
\makeatother

\makeatletter

\def\env@blscases{
  \let\@ifnextchar\new@ifnextchar
  \left.
  \def\arraystretch{1.2}
  \array{@{}l@{\quad}l@{}}
}
\makeatother

\makeatletter

\def\env@rcases{
  \let\@ifnextchar\new@ifnextchar
  \left.
  \def\arraystretch{1.2}
  \array{@{}l@{\quad}l@{}}
}
\makeatother

\hypersetup{%
    pdfsubject=Paper,
    pdfkeywords={nuclear physics} {MBPT} {dilute Fermi gas} {renormalization} {chiral EFT} {nuclear matter}
    unicode=true, 
    breaklinks=true,
     colorlinks   = true,
     linkcolor = blue,
     citecolor = blue,
     menucolor = blue,
     citecolor    = blue,
     urlcolor = blue
}

\allowdisplaybreaks
\begin {document}

\title{Dilute Fermi gas at fourth order in effective field theory}
\author{C. Wellenhofer}
\email[E-mail:~]{wellenhofer@theorie.ikp.physik.tu-darmstadt.de}
\affiliation{Institut f\"{u}r Kernphysik, Technische Universit\"{a}t Darmstadt, 64289 Darmstadt, Germany}  
\affiliation{ExtreMe Matter Institute EMMI, GSI Helmholtzzentrum f\"{u}r Schwerionenforschung GmbH, 64291 Darmstadt, Germany}  

\author{C. Drischler}
\email[E-mail:~]{cdrischler@berkeley.edu}
\affiliation{Department of Physics, University of California, Berkeley, CA 94720, United States of America}  
\affiliation{Lawrence Berkeley National Laboratory, Berkeley, CA 94720, United States of America}  

\author{A. Schwenk}
\email[E-mail:~]{schwenk@physik.tu-darmstadt.de}
\affiliation{Institut f\"{u}r Kernphysik, Technische Universit\"{a}t Darmstadt, 64289 Darmstadt, Germany}  
\affiliation{ExtreMe Matter Institute EMMI, GSI Helmholtzzentrum f\"{u}r Schwerionenforschung GmbH, 64291 Darmstadt, Germany}  
\affiliation{Max-Planck-Institut f\"{u}r Kernphysik, Saupfercheckweg 1, 69117 Heidelberg, Germany}  

\begin{abstract}
Using effective field theory methods, we calculate for the first time the complete fourth-order term in the Fermi-momentum or $k_{\rm F} a_s$ expansion for the ground-state energy of a dilute Fermi gas. The convergence behavior of the expansion is examined for the case of spin one-half fermions and compared against quantum Monte-Carlo results, showing that the Fermi-momentum expansion is well-converged at this order for $| k_{\rm F} a_s | \lesssim 0.5$.
\end{abstract}

\maketitle

The dilute Fermi gas has been a central problem for many-body calculations
for decades~\cite{Lenz,PhysRev.105.1119,1957PhRv,Efimov_1965,AMUSIA2,1965PhRv,Efimov2,AMUSIA1968377,RevModPhys.43.479,BISHOP1973106,PhysRevA.71.053605}.
Renewed interest in this problem has been triggered by striking progress with ultracold atomic gases.
In particular, 
by employing so-called Feshbach resonances~\cite{RevModPhys.82.1225} one can tune 
inter-atomic interactions and thereby probe Fermi systems over a wide range of many-body dynamics~\cite{RevModPhys.80.885}.
On the theoretical side, a systematic approach towards the dynamics of fermions (or bosons) at low energies has
emerged in the form of effective field theory (EFT)~\cite{Hammer:2000xg,Steele:2000qt,Furnstahl:2000we,Furnstahl:2001xq,Schafer:2005kg,Hammer:2016xye}.
Motivated by this,
we revisit
the expansion in the Fermi momentum $\kF$ of the ground-state energy density $E(\kF)$ of a dilute gas of one species of interacting fermions.
Using perturbative EFT methods, we calculate $E(\kF)$ up to fourth order in the expansion, including for the first time the 
complete fourth-order term.
From this, we analyze the convergence behavior of the expansion, 
and obtain precise predictions for $E(\kF)$ with systematic uncertainty estimates.
Our analytic results have important applications for various problems in many-body physics, including benchmarks for experimental and theoretical studies of cold atoms, the construction of improved models of neutron star crusts, and for
constraining nuclear many-body calculations at low densities.

Short-ranged EFT represents a systematic framework for the dynamics of fermions (or bosons) at low momenta $Q<\Lambda_b$, where $\Lambda_b$ denotes the breakdown scale.
At low momenta, details of the underlying interactions are not resolved and can be replaced by a series of contact interactions.
Few- and many-body observables are then expressed in terms of a systematic expansion in $Q/\Lambda_b$ (called ``power counting''). The EFT Lagrangian
is given by the most general operators consistent 
with Galilean invariance, parity, and time-reversal invariance. 
The low-energy constants of the Lagrangian have to be 
fitted to experimental data or (if possible) can be matched to the underlying theory.
Assuming spin-independent interactions, the (unrenormalized) Lagrangian reads (see, e.g., Refs.~\cite{Steele:2000qt,Hammer:2000xg,Furnstahl:2000we,Furnstahl:2001xq,Schafer:2005kg,Hammer:2016xye})
\begin{align} \label{Lagrangian} 
\mathscr{L}_\text{EFT} &= \psi^\dagger \left[i\partial_t +\frac{\ter \nabla^2}{2M} \right] \psi
-\frac{C_0}{2} (\psi^\dagger \psi)^2  
\breq
&\quad+\frac{C_2}{16} \left[(\psi \psi)^\dagger (\psi \te{\nabla}^2  \psi)+ \text{h.c.} \right]
\breq
&\quad+\frac{C'_2}{8} (\psi \te{\nabla}\psi)^\dagger \cdot(\psi \te{\nabla}\psi) 
- \frac{D_0}{6} (\psi^\dagger \psi)^3+ \ldots,
\end{align}
where $\psi$ are nonrelativistic fermion fields, $\te{\nabla}=\tel \nabla-\ter \nabla$ is the Galilean invariant derivative, h.c. the Hermitian conjugate,
and $M$ the fermion mass.

The ultraviolet (UV) divergences that appear beyond tree level in perturbation theory 
can be regularized by introducing a cutoff $\Lambda$
for relative momenta $\bs{p^{(\prime)}}$ and Jacobi momenta $\bs{q}^{(\prime)}$.
The two- and three-body potentials emerging from $\mathscr{L}_\text{EFT}$ are then given by 
\begin{align} \label{NNpotentials}
\braket{\bs{p'}|V^{(2)}_\text{EFT}|\bs{p}} 
&= \Big[ C_0(\Lambda) + C_2(\Lambda) (\bs{p'}^2+\bs{p}^2)/2 
\breq
&\quad+ C_2'(\Lambda)\,\bs{p'}\cdot\bs{p}\,+ \ldots\Big] 
\breq 
&\quad\times \theta(\Lambda-p)\theta(\Lambda-p'),
\\
\braket{\bs{p'}\bs{q'}|V^{(3)}_\text{EFT}|\bs{p}\bs{q}}
&=
\Big[D_0(\Lambda)+\ldots\Big]
\times \theta(\Lambda-p)\theta(\Lambda-q)\breq
&\quad \times 
\theta(\Lambda-p')\theta(\Lambda-q').
\end{align}
Perturbative renormalization is carried out by 
introducing counterterms such that the divergent contributions are canceled.
In the two-body sector, this leads to
\begin{align} \label{couplings1}
C_0(\Lambda)
&= C_0 + C_0 \sum_{\nu= 1}^3 \left(C_0\frac{M}{2\pi^2}\Lambda\right)^{\!\nu}
+C_2 C_0 \frac{M}{3\pi^2} \Lambda^3 + \ldots,
\\\label{couplings2}
C_2(\Lambda) 
&= C_2
+C_2 C_0 \frac{M}{\pi^2}\Lambda + \ldots,
\\\label{couplings3}
C_2'(\Lambda)
&= C_2'+\ldots ,
\end{align}
where
the cutoff-dependent parts
are counterterms.
For the renormalized two-body potential the residual cutoff dependence due to terms $\mathcal{O}(1/\Lambda)$ in perturbation theory vanishes in the limit $\Lambda\rightarrow \infty$. Matching the two-body low-energy constants to the effective-range expansion (ERE) then leads 
to (see, e.g., Ref.~\cite{Hammer:2000xg})
\begin{align} \label{LECs}
C_0 = \frac{4\pi a_s}{M}, \hspace*{8mm} C_2=C_0\frac{a_sr_{\!s}}{2}, \hspace*{8mm} C_2'=\frac{4\pi a_p^3}{M},
\end{align}
where $a_s$ and $a_p$ is the $S$- and $P$-wave scattering length, respectively, and $r_s$ is the $S$-wave effective range.

In the so-called natural case the low-energy constants scale according to
\begin{align}
C_0 \sim \frac{1}{M\Lambda_b}, \hspace*{9mm}
C_2 \sim C_2'\sim \frac{1}{M\Lambda_b^3},
\end{align}
so in this case low-energy observables can be calculated systematically by ordering
contributions in perturbation theory with respect to powers of $Q/\Lambda_b$.

In the two-body sector there are only
power divergences, but in systems with more than two particles also logarithmic divergences can occur, starting at order $(Q/\Lambda_b)^4$.
The counterterm for the leading logarithmic divergences is provided by the leading term of the three-body potential $V^{(3)}_\text{EFT}$.
Neglecting $\mathcal{O}(1/\Lambda)$ terms,
cutoff independence in the $N$-body sector with $N\geqslant 3$ at order  $(Q/\Lambda_b)^4$ is tantamount to
\begin{align} \label{RGE}
\frac{\partial}{\partial\Lambda} 
\left[
-(C_0)^4 \beta \ln\Lambda + D_0(\Lambda) \right]=0.
\end{align}
The coefficient of the $\ln\Lambda$ term in Eq.~\eqref{RGE} is 
$\beta=M^3( 4 \pi - 3\sqrt{3}\,)/(4\pi^3)$,
which can be obtained from the UV analysis of the two
logarithmically divergent three-body scattering diagrams at order $(Q/\Lambda_b)^4$, see Refs.~\cite{Efimov2,Braaten:1996rq,Hammer:2000xg}.
Integrating Eq.~\eqref{RGE}
leads to
\begin{align} 
D_0(\Lambda) 
=
D_0(\Lambda_0) + (C_0)^4  \beta \ln(\Lambda/\Lambda_0).
\end{align}
The low-energy constant $D_0(\Lambda_0)$
has to be fixed by matching to few-body data. For $\Lambda_0\sim \Lambda_b$ it is $D_0(\Lambda_0)\sim 1/(M\Lambda_b^4)$ in the natural case~\cite{Braaten1999}.
The scale $\Lambda_0$ is however completely arbitrary, with $D_0(\Lambda_0')=D_0(\Lambda_0)+(C_0)^4 \beta \ln(\Lambda_0'/\Lambda_0)$.

Applying the EFT potential $V_\text{EFT}=V_\text{EFT}^{(2)}+V_\text{EFT}^{(3)}$
in many-body perturbation theory (MBPT) leads 
to the Fermi-momentum expansion for the ground-state energy density $E(\kF)$ of the dilute Fermi gas, i.e.,
\begin{align} \label{kfexp0}
E(\kF) = n \frac{\kF^2}{2 M} \bigg[ \frac{3}{5} + (g-1) \sum_{\nu=1}^\infty \mathcal{C}_{\nu}(\kF)  \bigg],
\end{align}
where $n=g\,  \kF^3 /(6 \pi^2)$ is the fermion number density and $g$ is the spin multiplicity.
The dependence of a given MBPT diagram on $g$ is obtained
by inserting
a factor 
$\delta_{\sigma_1,\sigma'_1}\delta_{\sigma_2,\sigma'_2}-\delta_{\sigma_1,\sigma'_2}\delta_{\sigma_2,\sigma'_1}$ for each vertex and summing over the 
spins $\sigma^{(\prime)}_1$, $\sigma^{(\prime)}_2$ of the in- and outgoing 
lines.
Each MBPT diagram contributes only to a given order in the Fermi-momentum expansion, as specified by the EFT power counting.
This is in contrast to pre-EFT approaches to the dilute Fermi gas~\cite{1965PhRv,Efimov2,AMUSIA1968377,RevModPhys.43.479,BISHOP1973106}, 
which are complicated by summations to all orders and expansions for each diagram.

The leading term in the expansion was first obtained by Lenz~\cite{Lenz} in 1929, and
the second-order term was calculated by Lee and Yang~\cite{PhysRev.105.1119} as well as de Dominicis and Martin~\cite{1957PhRv} in 1957. 
They are given by
\begin{align}\label{E1}
\mathcal{C}_1(\kF) &= \frac{2}{3\pi} \kF a_s,
\\\label{E2}
\mathcal{C}_2(\kF) &= \frac{4}{35 \pi^2} (11- 2\ln 2) (\kF a_s)^2 .
\end{align}
The third-order term was first computed by 
de Dominicis and Martin~\cite{1957PhRv} in 1957
for hard spheres with two isospin states, by
Amusia and Efimov~\cite{AMUSIA2} in 1965 for a single species of hard spheres, and then 
by
Efimov~\cite{Efimov2} in 1966 for the general dilute Fermi gas.
It was also computed subsequently by various authors~\cite{AMUSIA1968377,RevModPhys.43.479,BISHOP1973106,PhysRevA.71.053605,Hammer:2000xg,Kaiser:2011cg,Kaiser:2012sr,Kaiser:2017xie}.
The most precise values have been obtained by Kaiser using semianalytic methods~\cite{Kaiser:2011cg,Kaiser:2012sr,Kaiser:2017xie}:
\begin{align}
\label{E3}
\mathcal{C}_3(\kF) &= \Big[0.0755732(0) +  0.0573879(0) \, (g-3) \Big](\kF a_s)^3
\breq
&\quad+\frac{1}{10\pi} (\kF a_s)^2 \kF r_{\!s} +  \frac{1}{5\pi}\frac{g+1}{g-1}(\kF a_p)^3 .
\end{align}
We have reproduced these results.
Our result for the
fourth-order term is given by
\begin{align} \label{E4}
\mathcal{C}_\text{4}(\kF) &= 
-0.0425(1)\,(\kF a_s)^4   +  0.0644872(0)\,  (\kF a_s)^3  \kF r_{\!s} 
\breq 
&\quad +\gamma_\text{4}\,(g-2)\, (\kF a_s)^4 ,
\end{align}
with  
\begin{align}  \label{gamma4}
\gamma_\text{4}(\kF)&= \frac{ M D_0(\Lambda_0)}{108\pi^4 a_s^4} +0.2707(4) - \, 0.00864(2)\,(g-2)
\breq
&\quad+\frac{16}{27 \pi^3}\left(4\pi-3\sqrt{3}\right)  \,
\ln(\kF/\Lambda_0).
\end{align}
Here, the effective-range contribution stems from
the two second-order diagrams with one $C_0$ and one $C_2$ vertex (plus the corresponding tree-level counterterm), which can be
evaluated using the semianalytic formula of Kaiser~\cite{Kaiser:2012sr}.
The remaining part of $\mathcal{C}_4(\kF)$ corresponds to diagrams with four $C_0$ vertices
and the tree-level contribution from $V_\text{EFT}^{(3)}$.

We note that Baker 
has published three different results for $\mathcal{C}_4(\kF)$ for $g=2$ in Refs.~\cite{1965PhRv,RevModPhys.43.479,Baker:1999np}.
In all of them, $\mathcal{C}_\text{4}(\kF)$ involves an additional parameter $A_0''$ that is presumed to be ``not determined by the two-body phase shifts''~\cite{RevModPhys.43.479,AMUSIA1968377}.
As is clear from the EFT perspective, the appearance of such a non-ERE parameter is not justified at this order (for $g=2$).
The first publication by Baker on the $\kF a_s$ expansion~\cite{1965PhRv} was criticized by Efimov and Amusia in Ref.~\cite{AMUSIA1968377}. 
Baker acknowledged this criticism and revised his result in Ref.~\cite{RevModPhys.43.479}. 
He later revised his $g=2$ result for $\mathcal{C}_4(\kF)/(\kF a_s)^4$ again in Ref.~\cite{Baker:1999np} (see Ref.~\cite{baker4}), where
he gives
for $r_s=a_p=0$ the value $-0.0372$, which is 
close to our $-0.0425(1)$. Note that we calculate
$\mathcal{C}_\text{4}(\kF)$ independently for both $g=2$ and for general $g$,
with $g\rightarrow 2$ matching the $g=2$ result.

Setting $\Lambda_0=1/|a_s|$ one obtains from the nonanalytic part of $\gamma_4(\kF)$ the known form of the logarithmic term at fourth order~\cite{Efimov_1965,Efimov2,AMUSIA1968377,RevModPhys.43.479,BISHOP1973106,Braaten:1996rq,Hammer:2000xg}.
Note again that $\Lambda_0$ is an arbitrary auxiliary scale: from Eq.~\eqref{RGE}, $\gamma_4(\kF)$ is independent of $\Lambda_0$.
Therefore, the logarithmic term
should not be treated as a separate contribution in the $\kF$ expansion.

\begin{table}
\caption
{Results for the regular contributions to $\mathcal{C}_\text{4}(\kF)$. Diagrams with $^{*}$ ($^{**}$) have UV power (logarithmic) divergences, which are subtracted by the respective counterterm contributions. 
Diagrams with $^{***}$ have infrared singularities. 
The uncertainty estimates take into account both the statistical Monte-Carlo uncertainties and variations of the cutoff. 
The $g$ factors are listed without the generic factor $g(g-1)$.}
\begin{center}
\begin{ruledtabular}
\begin{tabular}{llll}
\addlinespace
diagram  & $g$ factor & value \\
\hline
I1$^{*}$                               & $1$        & $+0.0383115(0)$ \\
I2$^{*}$+I3+I4$^{*}$+I5$^{*}$          & $1$        & $+0.0148549(0)$ \\
I6                                     & $1$        & $-0.0006851(0)$ \\
IA1                                    & $g(g-3)+4$ & $-0.003623(1)$ \\
%\hline
IA2                                    & $g(g-3)+4$ & $-0.001672(1)$ \\
IA3                                    & $g(g-3)+4$ & $-0.003343(1)$ \\
II1$^{*}$+II2$^{*}$                    & $g-3$      & $+0.058359(1)$ \\
II3+II4                                & $g-3$      & $-0.003358(1)$ \\
II5$^{**}$                             & $g-3$      & $+0.0645(1)$ \\
II6$^{**,*}$                           & $g-3$      & $-0.0265(2)$ \\
II7+II12                               & $g-3$      & $+0.003923(1)$ \\
II8+II11                               & $g-3$      & $+0.007667(1)$ \\
II9                                    & $g-3$      & $-0.000981(1)$ \\
II10                                   & $g-3$      & $-0.000347(1)$ \\
IIA1$^{**}$                            & $3g-5$     & $+0.0647(1)$ \\
IIA2+IIA4                              & $3g-5$     & $+0.004122(1)$ \\
IIA3                                   & $3g-5$     & $-0.000461(1)$ \\
IIA5                                   & $3g-5$     & $+0.003542(1)$ \\
IIA6                                   & $3g-5$     & $+0.003331(1)$ \\
III1$^{* {**},**,*}$+III7+III8$^{*{**},*}$  & $g-1$      & $-0.0513(2)$ \\
III2$^{*{**}}$+III9+III10$^{*{**}}$        & $g-1$      & $+0.001650(1)$ \\
\hline
(II5+IIA1)$_{g=2}$                       & $1$      & $+0.00018(1)$ \\
(II6+III1+III7+III8)$^{*}_{g=2}$             & $1$  & $-0.0248(1)$ \\
\hline
$\sum_{\text{diagrams}, g=2}$ &  $1$ & $-0.0425(1)$
\end{tabular}%
\end{ruledtabular}
\end{center}
\label{ourtable}
\end{table}

\begin{figure}[t]
\begin{center}
\includegraphics[width=0.425\textwidth]{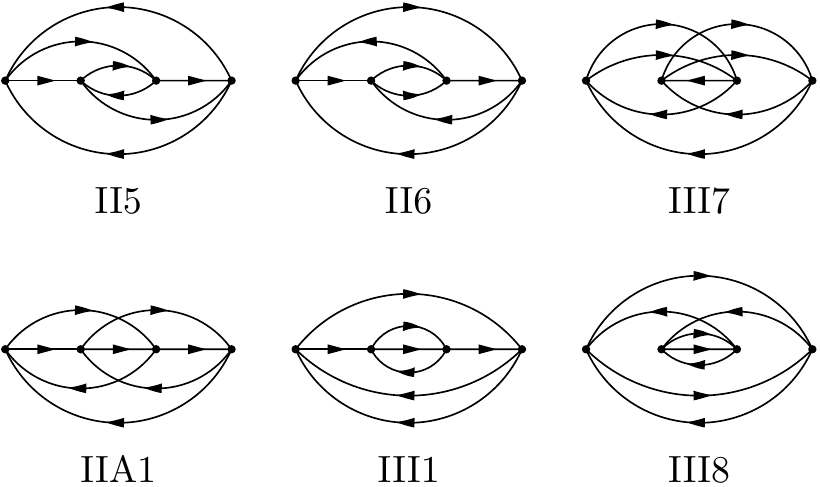} 
\end{center}
\caption{Hugenholtz diagrams representing the fourth-order contributions with logarithmic divergences II(5,6), IIA1, and III1.
Also shown are the other diagrams that are part of the sum III(1+7+8).
}% Holes have arrows pointed to the left, particles to the right.}
\label{MBPTdiags}
\end{figure}

For 
a momentum-independent 
potential 
(i.e., for the $C_0(\Lambda)$ part of $V_\text{EFT}^{(2)}$), only diagrams without 
single-vertex loops contribute at zero temperature.
There are 39 such diagrams at fourth order in MBPT~\cite{szabo,RevModPhys.43.479}, which
can be divided into four topological species:
\begin{itemize}
\item I(1-6): ladder diagrams,
\item IA(1-3): ring diagrams,
\item II(1-12), IIA(1-6): other two-particle irreducible diagrams,
\item III(1-12): two-particle reducible diagrams.
\end{itemize}
Here, we have  followed Baker's~\cite{RevModPhys.43.479} convention for the labeling of these diagrams according to groups that are closed under vertex permutations. 
Diagrams III(3,6,11,12) are anomalous and thus give no contribution in zero-temperature MBPT~\cite{Kohn:1960zz}.
The remaining diagrams are listed in Table~\ref{MBPTdiags}.
The following diagrams involve divergences:
\begin{itemize}
\item I(1,2,4,5), II(1,2,6), III(1,8): UV power divergences,
\item II(5,6), IIA1, III1: logarithmic UV divergences,
\item III(1,2,8,10): infrared divergences.
\end{itemize}
The UV divergences are removed by renormalization; i.e.,
the UV power divergences, which correspond to particle-particle ladders, are 
canceled by the counterterm contributions from the first-, second-, and third-order diagrams obtained by removing the ladders.
The diagrams with logarithmic divergences II(5,6), IIA1 and III1 are shown in Fig.~\ref{MBPTdiags}.
Using dimensionless momenta $\bs{i}\equiv \bs{k}_i/(\alpha\kF)$ one can analytically extract (in the limit $\Lambda\rightarrow\infty$) from each diagram a contribution 
$\sim \ln(\Lambda/(\alpha\kF))$. The parameter $\alpha$ is arbitrary, and can be set to $\alpha=1$. Adding the 
logarithmic part of the  tree-level contribution from $V_\text{EFT}^{(3)}$, this
leads to the logarithmic part of $\mathcal{C}_4(\kF)$ given in Eq.~\eqref{gamma4}.
Finally, the infrared divergences are due to repeated energy denominators. 
This is a generic feature of two-particle reducible contributions in zero-temperature MBPT 
(see also~\cite{RevModPhys.43.479}, Sec. III.C., and~\cite{Feldman1996}, Sec. 1.4.).
At each order, the infrared singularities are removed when certain two-particle reducible diagrams are combined, in the present case III(1+8) and III(2+10).
More details on the calculation of the fourth-order MBPT diagrams are given in the appendix.
We have carried out the numerical calculations using the Monte-Carlo framework introduced in Ref.~\cite{Drischler:2017wtt} to evaluate high-order many-body diagrams.

\begin{figure*}[t]
\begin{center}
\includegraphics[width=\textwidth]{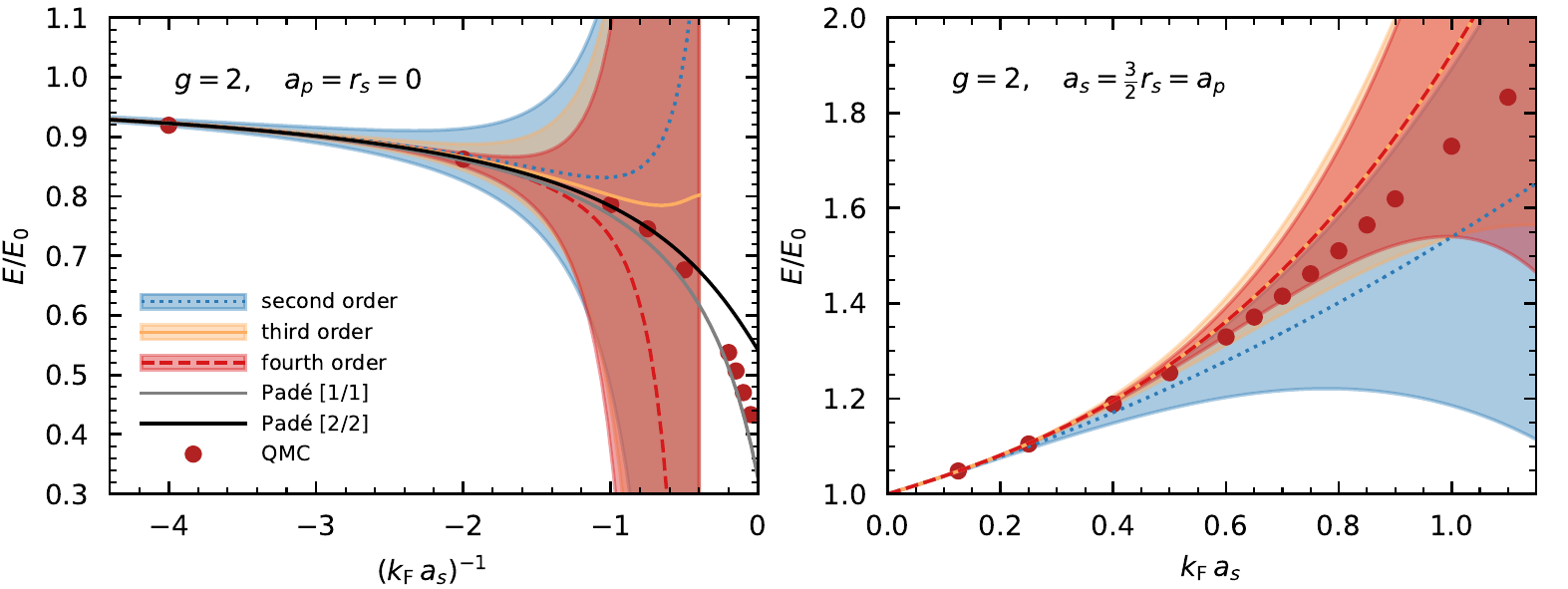} 
\end{center}
\vspace*{-2mm}
\caption{Results for $E/E_0$ from the Fermi-momentum expansion and from QMC calculations~\cite{Gandolfi:2015jma,Gandolfi,Pila10FerroQMC}, see text for details.
For clarity, in the left panel the order-by-order results are plotted only up to $\kF a_s = -2.5$. Note the systematic order-by-order improvement with overlapping 
uncertainty bands.}
\label{plot}
\end{figure*} 

Our results for the various contributions to the regular (i.e., nonlogarithmic) part of $\mathcal{C}_4(\kF)$ are listed in Table~\ref{MBPTdiags}.
The numerical values for the diagrams without divergences are similar (but small differences are present) to the ones published by 
Baker in Table~IV of Ref.~\cite{RevModPhys.43.479}.
The contributions that involve logarithmic divergences, II5, II6, IIA1, and III(1+7+8), have the largest numerical uncertainties.
For $g=2$ slightly more precise results can be given for II5+IIA1 and II6+III(1+7+8), because then no logarithmic divergences occur.

For spin one-half fermions, the logarithmic term at fourth order (and beyond, up to a certain order $N_\text{log}$) is Pauli blocked, so in that case the 
$\kF$ expansion is (for $N<N_\text{log}$) given by
\begin{align} \label{EE0plot}
g=2:\;\;\;\;\;E(\kF)= E_0 \left(
1+\sum_{\nu=1}^N X_\nu \delta^\nu \right)+ o(\delta^{N}),
\end{align}
where $\delta=\kF a_s$ and $E_0 = 3 n\kF^2/(10 M)$. The coefficients $X_\nu$ are completely determined by the ERE parameters.
For $r_s=a_p=0$ (LO), the coefficients are 
\begin{align} \label{EE0plotcoeffs}
(X_1,X_2,X_3,X_4)=
(+0.354,
+0.186,
+0.030,
-0.071),
\end{align}
and for the hard-sphere gas (HS) with $a_s=3r_s/2=a_p$, we obtain
\begin{align} \label{EE0plot2}
(X_1,X_2,X_3,X_4)=
(+0.354,
+0.186,
+0.384,
+0.001).
\end{align}
The results for $N\in\{2,3,4\}$ are plotted in Fig.~\ref{plot}. For comparison, we also
show results obtained from quantum Monte-Carlo (QMC) calculations~\cite{Gandolfi:2015jma,Gandolfi,Pila10FerroQMC}.
Overall, the perturbative results are very close to the QMC results for $|\delta| \lesssim 0.5$ and start to deviate strongly for $|\delta| \gtrsim 1$.
In the LO case, 
the relative error with respect to the QMC point at $\delta = - 0.5$ is $4.5\%$ at first,
$0.8\%$ at second, $0.4\%$ at third, and $0.1\%$ at fourth order. 
In the HS case $X_4$ is very small and the $N=3$ and $N=4$ curves are almost indistinguishable.

In Fig.~\ref{plot} we 
also plot uncertainty bands
obtained by setting
$X_{N+1} = \pm \max[X_{\nu\leq N}]$. 
Going to higher orders in that scheme reduces the width of the bands in the perturbative region $|\delta| \lesssim 1$.
For $|\delta| \lesssim 0.5$ the bands are very small for $N=4$, which supports the conclusion 
that the expansion is well-converged at fourth order in this regime.
Note that these results do not depend on $a_s$ being of natural size; only $\kF a_s$ has to be small.

For the case where $a_s$ is large, 
resummation methods
provide a means to extrapolate to larger values of $\kF a_s$.
One possible method, which was employed also by Baker~\cite{Baker:1999np,baker4}, is to use Pad{\'e} approximants~\cite{padenote,bakerbook}.
The LO results obtained from the Pad{\'e}~$[1,1]$ and $[2,2]$ approximants are plotted in Fig.~\ref{plot}. Only diagonal Pad{\'e} approximants have a meaningful unitary limit.
The Pad{\'e}~$[2,2]$ results are very close to the QMC points for $\delta \lesssim -1.2$, while the Pad{\'e}~$[1,1]$ ones are in better agreement
with the QMC points close to the unitary limit $\delta\rightarrow -\infty$.
Note that pairing effects, which become relevant for
larger values of $-\delta$, can be expected 
to influence the large-order behavior of the Fermi-momentum 
expansion~\cite{Marino2019}.
The range for the Bertsch parameter obtained from the Pad{\'e}~$[1,1]$ and $[2,2]$ approximants, $\xi_\text{Pad{\'e}}\in[0.33,0.54]$, is consistent 
with the value $\xi\approx 0.376$ extracted from experiments with cold atomic gases, and also with
the extrapolated value for the normal (i.e., non-superfluid) Bertsch parameter $\xi_n\approx 0.45$~\cite{Ku563}. 
Altogether, these
results may indicate that Pad{\'e} approximants converge in a larger region, compared to the Fermi-momentum expansion.
To further investigate this one would need to
construct the subsequent Pad{\'e}~$[\nu,\nu]$ approximants, which require the expansion coefficients up to order $2\nu\geqslant 6$.

In summary, 
using EFT methods we have calculated the complete fourth-order term in the Fermi-momentum expansion for the ground-state energy of a dilute Fermi gas.
A detailed study of the convergence behavior and comparison against QMC calculations for the case of spin one-half fermions showed that this (asymptotic) expansion is well-converged at this order for $|\kF a_s| \lesssim 0.5$, and exhibits divergent behavior for $|\kF a_s| \gtrsim 1$.
Our results provide important high-order benchmarks for many problems in many-body physics, ranging from cold atomic gases to dilute nuclear matter and neutron stars.
\\
\\
\noindent
\textbf{Acknowledgements}
\\

We thank R.F.~Bishop, R.J.~Furnstahl, A.~Gezerlis, K. Hebeler, S.~K{\"o}nig, K.~McElvain, D.~Phillips and A.~Tichai for useful discussions, and S.~Gandolfi as well as \mbox{S.~Pilati} for sending us their QMC results. This work is supported in part by  the  Deutsche Forschungsgemeinschaft (DFG, German Research Foundation) -- \mbox{Projektnummer} 279384907 -- SFB 1245, the US Department of Energy, the Office of Science, the Office  of  Nuclear  Physics,  and  SciDAC  under awards \mbox{DE-SC00046548} and \mbox{DE-AC02-05CH11231}. C.D. acknowledges  support by the  Alexander von Humboldt Foundation through a Feodor-Lynen Fellowship. Computational resources have been provided by the Lichtenberg high performance computer of the TU Darmstadt.
\\
\\
\noindent
\textbf{Appendix}
\\
\appendix
Here, we provide more details regarding the evaluation of the fourth-order MBPT diagrams.

The diagrams in the pairs I(3,4), III(7,8) and III(9,10) can be combined to get simplified energy denominators;
I(2,5), II(1,2), II(3,4), II(7,8), II(11,12) and IIA(2,4) give identical results for a spin-independent potential; and
for a momentum-independent potential the contribution from I(3+4) is half of that from I(2+5).
The diagrams I(1-6) can be calculated using the semianalytic expressions derived by Kaiser~\cite{Kaiser:2011cg}, which can be obtained from the usual MBPT expressions~\cite{szabo} by applying various partial-fraction decompositions and the Poincar\'{e}-Bertrand transformation formula~\cite{Muskheli}.
For the numerical evaluation of the IA diagrams it is more convenient to use single-particle momenta instead of relative momenta, because then the 
phase space is less complicated.
The II, IIA and III diagrams without divergences can be evaluated in the same way as the IA diagrams.

The expression for III(1+7+8) is given by
\begin{align} \label{III178}
E_\text{4,III(1+7+8)}&=
-\zeta (g-1) \sum_{ \substack{\bs{i},\bs{j},\bs{k} \\ \bs{a},\bs{c} }  }
n_{ijk} \bar n_{abc} \frac{\theta_{\bs{ab}}}{\mathcal{D}_{ab,ij}^2 } 
\breq
&\quad\times  
\left( \bar n_d\frac{\theta_{\bs{ka}}\theta_{\bs{cd}}}
{\mathcal{D}_{bcd,ijk}} 
- \bar n_{d'} \frac{\theta_{\bs{cd'}}}
{\mathcal{D}_{cd',ik}} \right)
\bigg|{ \substack{  {\color{white}dummy} \\\bs{b}=\bs{i}+\bs{j}-\bs{a}\\ \bs{d}=\bs{k}+\bs{a}-\bs{c}\\ \bs{d'}=\bs{i}+\bs{k}-\bs{c} }}\;,
\end{align}
Here, $\sum_{\bs{i}}\equiv \int d^3 i/(2\pi)^3$, the distribution functions are $n_{ij\ldots}\equiv n_i n_j \cdots$ and
$\bar n_{ab\ldots}\equiv \bar n_a \bar n_b \cdots$, with 
$n_i\equiv\theta(1-i)$ and $\bar n_a\equiv\theta(a-1)$, and
the energy denominators are given by $\mathcal{D}_{ab,ij}\equiv(a^2+b^2-i^2-j^2)/(2M)$. 
Moreover, $\zeta=\kF^9  g(g-1)(C_0)^4$
and $\theta_{\bs{ab}}\equiv \theta(\Lambda/\kF-|\bs{a}-\bs{b}|/2)$.
For details on the diagrammatic rules, see, e.g., Ref.~\cite{szabo}.
The infrared divergence corresponds to $\mathcal{D}_{ab,ij}=0$, and in that case the two terms in the large brackets cancel each other, and similar for III(2+10).
For III(1+8) also 
the linear UV divergences are removed (the counterms for the power divergences of III1 and III8 would come from diagrams with single-vertex loops).
The remaining logarithmic UV divergence is given by
\begin{align}
\frac{E_\text{4,III(1+7+8)}}{\ln(\Lambda/\kF)}\xrightarrow{\Lambda\rightarrow \infty}
\zeta(g-1)\frac{\sqrt{3}}{3^3 2^7 \pi^{9}} .
\end{align}
Subtracting this term from Eq.~\eqref{III178} enables the numerical evaluation of the regular (i.e., nonlogarithmic) contribution from III(1+7+8) to $\mathcal{C}_4(\kF)$. 
The evaluation of the regular contributions from II5 and IIA1 is similar, i.e., the corresponding $\ln(\Lambda/\kF)$ terms have to be subtracted.

This leaves the diagrams with power divergences II(1,2,6), where diagram II6 has also a logarithmic divergence.
The expression for II6 reads
\begin{align} \label{II6}
E_\text{4,II6}&=
-  \zeta(g-3)   \sum_{ \substack{\bs{i},\bs{j},\bs{k} \\ \bs{a},\bs{c}}  }  n_{ijk} \bar n_{abcde} 
\theta_{{\bs{a}\bs{b}}} \theta_{{\bs{k}\bs{a}}} \theta_{{\bs{c}\bs{d}}} \theta_{{\bs{j}\bs{e}}} \theta_{{\bs{b}\bs{e}}}
\breq
&\quad\times
\frac{1}
{\mathcal{D}_{ab,ij}\mathcal{D}_{be,ik} \mathcal{D}_{bcd,ijk}} 
\bigg|{\substack{ {\color{white}dummy} \\ \bs{b}=\bs{i}+\bs{j}-\bs{a}\\ \bs{d}=\bs{k}+\bs{a}-\bs{c} \\ \bs{e}=\bs{k}+\bs{a}-\bs{j}}}\;.
\end{align}
Here, $\theta_{{\bs{k}\bs{a}}}$,
$\theta_{{\bs{j}\bs{e}}}$ and $\theta_{{\bs{b}\bs{e}}}$ are redundant.
Substituting $\bs{K}=(\bs{i}+\bs{j})/2$, $\bs{p}=(\bs{i}-\bs{j})/2$, $\bs{z}=\bs{k}$, $\bs{A}=(\bs{a}-\bs{b})/2$, 
and $\bs{Y}=(\bs{c}-\bs{d})/2$ leads to
\begin{align} \label{II6b}
E_\text{4,II6}&=- 8M^3\,\zeta(g-3)   \sum_{ \substack{\bs{K},\bs{p},\bs{z} \\ \bs{A},\bs{Y}}}n_{ijk}  \bar n_{abcde}
\, \theta_A  \theta_Y\,\frac{1 }{A^2-p^2}
\breq
&\quad\times \frac{1 }
{\big[(\bs{A}+\bs{p})\cdot(\bs{A}-\bs{K}+\bs{z})\big] (Y^2-p^2+\mathcal{R})},
\end{align}
where $\mathcal{R}=(3\bs{A}+\bs{K}-\bs{z})\cdot(\bs{A}-\bs{K}+\bs{z})/4$ and $\theta_A\equiv \theta(\Lambda/\kF-A)$.
The two divergences of II6 can now be separated via
\begin{align} \label{dec}
\frac{1 }{Y^2-p^2+\mathcal{R}} = 
\underbrace{\frac{1}{Y^2}}_{\leadsto E_\text{4,II6(i)}}
+
\underbrace{ \frac{p^2-\mathcal{R}}{(Y^2-p^2+\mathcal{R}) Y^2}}_{\leadsto E_\text{4,II6(ii)}},
\end{align}
with $E_\text{4,II6(i)} \sim \Lambda$ for $\Lambda\rightarrow \infty$, and 
\begin{align}
\frac{E_\text{4,III6(ii)}}{ \ln(\Lambda/\kF)}
\xrightarrow{\Lambda\rightarrow \infty}
\zeta(g-3)\frac{\sqrt{3}}{3^3 2^7 \pi^{9}}.
\end{align}
The evaluation of the contribution from III6(ii) is similar to III(1,7,8), II5, and IIA1.
For III6(i), the effect of the counterterm can be implemented via the identity
\begin{align} \label{dec2}
\frac{\Lambda}{2\pi^2}-\sum_{\bs{Y}}\bar n_{cd}\frac{\theta_Y}{Y^2}  
\xrightarrow{\Lambda\rightarrow\infty}
\sum_{\bs{Y}}(n_c+n_d-n_{cd})\frac{\theta_Y}{Y^2}.
\end{align}
For diagrams II(1,2) as well as I(1,2,4,5), the same procedure can be applied.
For I(1,2,4,5) we have reproduced the semianalytic results in this way.

%------------------------------------------------------------
\bibliographystyle{apsrev4-1}		
\bibliography{refs}		

\begin{thebibliography}{38}
\bibitem [{pad()}]{padenote}%
  \BibitemOpen
  \href@noop {} {}\bibinfo {howpublished} {For a given formal power series
  $f(x)\simeq \sum_{k=0}^\infty f_k x^k$, the Pad{\'e} $[n,m]$ approximant is
  the rational function $f_{[n,m]}(x)=\frac{\sum_{k=0}^n {a_k}
  x^k}{1+\sum_{l=1}^m b_l x^l}$ whose Maclaurin expansion matches the series up
  to order $x^{n+m}$.}\BibitemShut {Stop}%
\end{thebibliography}
%-----------------------------------------------------------------

\end{document}